\documentclass[preprint,12pt]{elsarticle}

\usepackage{amssymb}
\usepackage{amsmath}
\usepackage{booktabs}
\usepackage{floatrow}
\usepackage{textcomp}
\usepackage[utf8]{inputenc}
\usepackage{graphicx}
\usepackage[version=4]{mhchem}
\usepackage{siunitx}
\usepackage{longtable,tabularx}
\usepackage{tikz}
\usepackage{algpseudocode}
\usepackage{algorithm}
\usepackage{xcolor}
\usepackage{enumitem}
\usepackage{doi}

\newenvironment{breakablealgorithm}
  {
   \begin{center}
     \refstepcounter{algorithm}
     \hrule height.8pt depth0pt \kern2pt
     \renewcommand{\caption}[2][\relax]{
       {\raggedright\textbf{\fname@algorithm~\thealgorithm} ##2\par}%
       \ifx\relax##1\relax 
         \addcontentsline{loa}{algorithm}{\protect\numberline{\thealgorithm}##2}%
       \else 
         \addcontentsline{loa}{algorithm}{\protect\numberline{\thealgorithm}##1}%
       \fi
       \kern2pt\hrule\kern2pt
     }
  }{
     \kern2pt\hrule\relax
   \end{center}
  }

\journal{Journal of Sound and Vibration}

\begin{document}

\begin{frontmatter}



\title{Neural network based approach for solving problems in plane wave duct acoustics}


\author[label1]{D. Veerababu}

\affiliation[label1]{organization={Department of Electrical Engineering},
            addressline={Indian Institute of Science}, 
            city={Bengaluru},
            postcode={560012}, 
            state={Karnataka},
            country={India}}

\author[label1]{Prasanta K. Ghosh\corref{cor1}}

\cortext[cor1]{Corresponding author.}
\ead{prasantg@iisc.ac.in}

\begin{abstract}
Neural networks have emerged as a tool for solving differential equations in many branches of engineering and science. But their progress in frequency domain acoustics is limited by the vanishing gradient problem that occurs at higher frequencies. This paper discusses a formulation that can address this issue. The problem of solving the governing differential equation along with the boundary conditions is posed as an unconstrained optimization problem. The acoustic field is approximated to the output of a neural network which is constructed in such a way that it always satisfies the boundary conditions. The applicability of the formulation is demonstrated on popular problems in plane wave acoustic theory. The predicted solution from the neural network formulation is compared with those obtained from the analytical solution. A good agreement is observed between the two solutions. The method of transfer learning to calculate the particle velocity from the existing acoustic pressure field is demonstrated with and without mean flow effects. The sensitivity of the training process to the choice of the activation function and the number of collocation points is studied. 

\end{abstract}



\begin{highlights}
\item PINNs-based solution for problems in frequency domain acoustics is presented.
\item Problems associated with the existing Lagrange multiplier method are discussed.
\item Acoustic field is predicted in the ducts of two different cross-sections.
\item Mean flow and visco-thermal effects are captured with a maximum error of $\mathcal{O}(10^{-4})$.
\item Results show the ability of PINNs to emerge as an acoustic solver in the future.
\end{highlights}


\end{frontmatter}



\newcommand{\blueline}{\raisebox{2pt}{\tikz{\draw[-,blue,solid,line width = 1.5pt](0,0) -- (10mm,0);}}}
\newcommand{\bluedotline}{\raisebox{2pt}{\tikz{\draw[-,blue,dotted,line width = 1.5pt](0,0) -- (10mm,0);}}}
\newcommand{\redline}{\raisebox{2pt}{\tikz{\draw[-,red,dashed,line width = 1.5pt](0,0) -- (10mm,0);}}}
\newcommand{\blackline}{\raisebox{2pt}{\tikz{\draw[-,black,solid,line width = 1.5pt](0,0) -- (10mm,0);}}}

\section{Introduction}\label{Sec:1}
Advancements in data-driven methodologies have revolutionized both science and engineering disciplines. Neural networks enabled researchers to extract valuable insights from large and complex datasets. They have been successfully used in structural health monitoring \cite{Wang2007,Janssens2016,Lai2021}, especially crack and damage detection \cite{Park2009,Abdeljaber2017}. In the area of acoustics, neural networks have been shown to be successful in capturing wave propagation phenomena from numerical simulation datasets. Stefan Sack and Mats {\AA}bom \cite{Sack2020} used neural networks to decompose plane waves inside a uniform duct. In their work, they used numerical simulation data obtained by solving Linearized Navier-Stokes equations. Alguacil et al. \cite{Alguacil2021} modeled sound propagation in a two-dimensional quiescent media with the help of neural networks using the database from Lattice-Boltzmann temporal simulations.  Although neural networks offer promising results with numerical simulation data (in a supervised way), in many other cases, the acquired data alone may not be sufficient to accurately model physical phenomena due to incomplete measurements, limited observations, or the presence of underlying physical laws \cite{de2020}. In such scenarios, the incorporation of prior knowledge and physical constraints becomes crucial to build reliable and interpretable models. 

By incorporating prior knowledge of the underlying physics and exploiting the expressive power of neural networks, physics-based neural network formulation provide a versatile framework for capturing complex relationships between dependent and independent variables in the governing equations as well as initial conditions (ICs) and/or boundary conditions (BCs) \cite{Raissi2019}. This mesh-less framework can offer reliable and faster solutions complementary to traditional mesh-based finite element methods \cite{Ihlenburg1995}, boundary element methods \cite{Shen2007,Van2021}, and computational fluid dynamics techniques \cite{Lourier2012}. In the neural network formulation, feedforward neural networks are used as function appproximators, and the problem of solving the governing differential equations along with the ICs and/or BCs is posed as an optimization problem. The function to be optimized, known as the loss function, is calculated as a weighted summation of the individual loss functions computed as the residuals of the governing differential equations, ICs and/or BCs upon approximating the desired output variables to a feedforward neural network \cite{Van2022,Maddu2022}. 

Several research groups successfully implemented neural network formulation to solve Burgers' equation \cite{Raissi2019}, Schr{\"o}ndinger equation \cite{Raissi2019}, Klein-Gordon equation \cite{Basir2022}, Poisson equation \cite{Basir2022}, etc. Attempts have been carried out to solve the Navier-Stokes equation as well \cite{Baymani2015,Oldenburg2022,Amalinadhi2022}. This article demonstrates the inherent flexibility of the feedforward neural network to efficiently represent complex wave behavior while satisfying the governing acoustic wave equations and the BCs in the frequency domain. Implementation of the neural network formulation for the Helmholtz equation can be evidenced in the literature. However, the analysis is limited to the lower frequencies \cite{Alkhalifah2021,Basir2022,Song2022,Wang2021}. The frequency-dependent behavior of the Helmholtz equation and its sensitivity to boundary conditions present significant challenges in obtaining accurate and efficient solutions, particularly in the high-frequency regimes. At higher frequencies, the network encounters \emph{vanishing gradient problem} which causes the training process to bias towards one of the loss functions \cite{Wang2021}. This results in inaccurate predictions. By tuning the individual weights associated with each loss function, the problem can be circumvented. However, these weights are sensitive to frequency and BCs. The weights chosen for a particular frequency and BCs are not suitable for others. Automatic weight update algorithms have been developed to dynamically adjust weights during each iteration or over a set of iterations \cite{Van2022,Maddu2022,Basir2022,Wang2021}. However, most of these algorithms require manual tuning of the hyperparameters which vary for each frequency and chosen BCs. In addition, prediction of complex-valued acoustic field due to sound propagation in absorptive medium or inside the narrow ducts is challenging as the loss function to be minimized has to be always a real-valued function. 

This article adopts a methodology that addresses the aforementioned difficulties and offers a reliable solution to the fundamental problems in the frequency domain acoustics. The versatility and efficacy of the methodology is demonstrated through comprehensive numerical experiments and comparisons with analytical solutions. The following three classical problems from the plane wave duct acoustic theory are chosen for demonstration purposes \cite{Munjal2014}. 
\begin{enumerate}[label=(\roman*)]
    \item Prediction of the acoustic field in a uniform duct \cite{Munjal2014}
    \item Prediction of the acoustic field in a gradually varying cross-sectional area duct \cite{Pillai2019}
    \item Prediction of complex-valued acoustic pressure in a narrow uniform duct \cite{Allard2009}
\end{enumerate}

In addition to training the network from scratch, the use of a \emph{transfer learning approach} \cite{Desai2021,Gao2022,Pellegrin2022,Xu2023} is demonstrated. In this approach, a new neural network is trained using the output of an existing neural network. This technique enables us to predict other acoustic variables such as particle velocity, acoustic impedance, power, intensity, etc., from the existing variables. In this work, the transfer learning technique is demonstrated by predicting the particle velocity from the existing acoustic pressure in a uniform duct with and without convective mean flow effects. The potential challenges and limitations associated with the application of neural networks in solving frequency domain acoustic problems are also discussed. These include network architecture resolution, especially training data, and the activation function to ensure robust and stable convergence. 

The article is organized as follows. Section~\ref{Sec:2} gives a theoretical description of the physics-based neural network formulation and its application to chosen problems. The results obtained from the neural network formulation are compared with the analytical solutions in Section~\ref{Sec:3}. The method of transfer learning and the sensitivity study of the neural network architecture to different frequencies are also discussed in the same section. The article is concluded in Section~\ref{Sec:4} with final remarks.


\section{Physics-based neural network formulation}\label{Sec:2}
Let us consider a problem of solving a differential equation 
\begin{equation}
    \mathcal{D}[\,\psi(\mathbf{x})\,] = s(\mathbf{x}), \qquad \forall\,\, \mathbf{x}\in\Omega, \label{Eq:1}
\end{equation}
subjected to the boundary conditions 
\begin{equation}
    \psi(\mathbf{x}) = g(\mathbf{x}), \qquad \forall\,\,\mathbf{x}\in \partial\Omega, \label{Eq:2}
\end{equation}
where $\mathcal{D}[\,\cdot \,]$ is the differential operator, $\psi(\mathbf{x})$ is the field variable, $s(\mathbf{x})$ is the forcing function, $g(\mathbf{x})$ is the set of prescribed boundary values, $\Omega$ is the domain, $\partial\Omega$ is the boundary of the domain, and $\mathbf{x}$ is the spatial vector. 

According to the universal approximation theorem (UAT), any field variable such as $\psi(\mathbf{x})$ considered in the above problem can be approximated by a neural network provided the field variable $\psi(\mathbf{x})$ is a continuous and bounded function within the domain of interest \cite{Hornik1989}. It is important to note here that the UAT states the possibility of finding the neural network with the desired accuracy. However, it does not provide sufficient information on the required network architecture and optimization algorithm to find it.

\begin{figure}[h!]
	\includegraphics[scale=0.9]{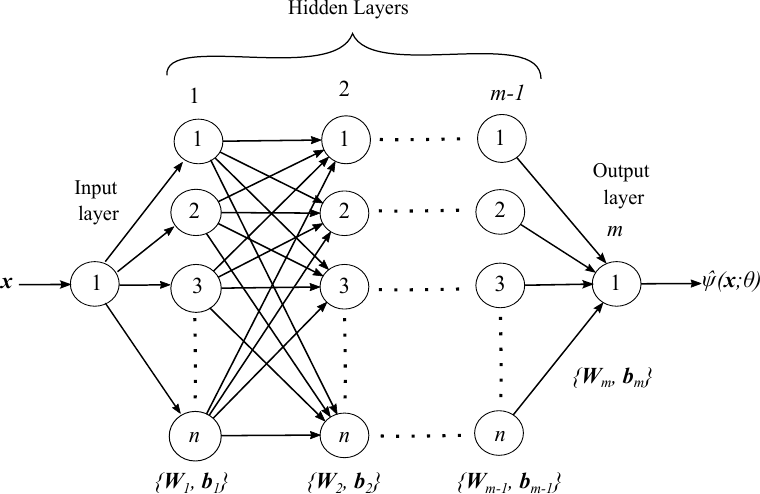}\centering
	\caption{\label{fig:1}{Schematic diagram of a feedforward neural network.}}
	\end{figure}

Figure~\ref{fig:1} shows the schematic diagram of a feedforward neural network with $m$ layers and $n$ neurons in each hidden layer.

A typical feedforward neural network used in the physics-based neural network formulation does not take the field variable information either from the experiments or from other numerical simulations as input. Instead, it takes the domain information ($\mathbf{x}$) in a descritized format. It is then passed from the input layer to the first hidden layer, where it undergoes through a nonlinear function along with the weights and biases associated with that particular neuron.

If $\mathbf{f}_q$ represents the output of the $q$-th hidden layer, where $q=$ 1, 2, 3, ..., $m-1$, then the feedforward network shown in Fig.~\ref{fig:1} is mathematically represented as follows \cite{Dung2023}
\begin{align}
    \mathbf{f}_0 &= \mathbf{x}, \\
    \mathbf{f}_q &= \sigma(\mathbf{W}_q\mathbf{f}_{q-1}+\mathbf{b}_q), \\
    \mathbf{f}_m &= \mathbf{W}_m\mathbf{f}_{m-1}+\mathbf{b}_m,
\end{align}
where $\mathbf{f}_0$ is the output of the input-layer, $\mathbf{f}_m$ is the output of the output-layer, $\left\{\mathbf{W}_q, \mathbf{b}_q\right\}$ and $\left\{\mathbf{W}_m, \mathbf{b}_m\right\}$ are the set of weights and biases of the $q$-th hidden layer and the output-layer, respectively. Here, $\sigma$ is called the activation function through which nonlinearity is introduced into the network. If $\hat{\psi}(\mathbf{x};\theta)$ represents the neural network approximation for $\psi(\mathbf{x})$, then $\hat{\psi}(\mathbf{x};\theta)=\mathbf{f}_m$, where $\theta=\left\{\mathbf{W}_q, \mathbf{b}_q, \mathbf{W}_m, \mathbf{b}_m\right\}$, $q=$ 1, 2, 3, ..., $m-1$, are the parameters of the network.

Now, $\hat{\psi}(\mathbf{x};\theta)$ can be found by solving the following optimisation problem \cite{Basir2022}
\begin{equation}
\begin{aligned}
\min_{\theta} \quad & \mathcal{L}_d(\mathbf{x}_d;\theta), \quad \mathbf{x}_d\in\Omega \\
\textrm{s.t.} \quad & \mathcal{L}_b(\mathbf{x}_b;\theta)=0, \quad \mathbf{x}_b\in\partial\Omega \label{Eq:6}
\end{aligned}
\end{equation}
where $\mathcal{L}_d$ and $\mathcal{L}_b$ are the loss functions associated with the differential equation and the boundary conditions, respectively. They are defined as follows
\begin{align}
    \mathcal{L}_d(\mathbf{x}_d;\theta) &= \frac{1}{N_d}\sum_{i=1}^{N_d}\left\|\mathcal{D}[\,\hat{\psi}(x^{(i)}_d;\theta)\,]-s(x^{(i)}_d)\right\|^2_2 \label{Eq:4}, \\
    \mathcal{L}_b(\mathbf{x}_b;\theta) &= \frac{1}{N_b}\sum_{i=1}^{N_b}\left\|\hat{\psi}(x^{(i)}_b;\theta)-g(x^{(i)}_b)\right\|^2_2. \label{Eq:8}
\end{align}
Here, $\left\|\,\cdot\,\right\|_2$ represents the $L_2$-norm, $N_d$ and $N_b$ are the number of collocation points inside the domain and on the boundary with $i$-th point is represented by $x^{(i)}_d$ and $x^{(i)}_b$, respectively.

It is a constrained optimization problem. It can be converted into an unconstrained optimisation problem using two approaches:\\ \\
1) \underline{Lagrange Multiplier Approach:} \\ \\
Using the Lagrange multiplier $\lambda_j$, the constrained optimisation problem in Eq.~(\ref{Eq:6}) can be converted into an unconstrained optimisation problem as follows \cite{Basir2022}
\begin{equation}
\min_{\theta} \quad \mathcal{L}_d+\sum_{j=1}^{M}\lambda_j\mathcal{L}_{b,j}, \label{Eq:9}
\end{equation}
where $j=$ 1, 2, 3, ..., $M$ are the boundaries with the corresponding loss functions $\mathcal{L}_{b,j}$ and the Lagrange multipliers $\lambda_j$. \newpage \noindent
2) \underline{Trial Solution Approach:} \\ \\
In this approach, a trial neural network is constructed that exactly satisfies the boundary conditions prior to training the network. Since the network already satisfies the boundary conditions, their contribution towards the total loss function will be dropped and the optimisation problem can be written as \cite{Lagaris1998}
\begin{equation}
\min_{\theta} \quad \mathcal{L}_d(\mathbf{x};\theta), \quad \mathbf{x}=\{\mathbf{x}_d,\mathbf{x}_b\}, \label{Eq:10}
\end{equation}
where
\begin{equation}
    \mathcal{L}_d(\mathbf{x};\theta) = \frac{1}{N}\sum_{i=1}^{N}\left\|\mathcal{D}[\,\hat{\psi}_t(x^{(i)};\theta)\,]-s(x^{(i)})\right\|^2_2. \label{Eq:11}
\end{equation}
Here, $\hat{\psi}_t(\mathbf{x};\theta)$ is the trial neural network, and $N$ is the number of collocation points of the entire domain including the boundaries. 

Both these approaches are equally efficient in solving a wide variety of governing equations occurring in many branches of engineering and physical sciences. However, in solving certain class of problems such as \emph{acoustic field prediction in the frequency domain}, the former approach encounters vanishing gradient problem at higher frequencies. That is, the gradients of the loss function $\mathcal{L}_b$ with respect to the parameters become almost equal to zero at higher frequencies. Due to this, the parametric update with respect to loss function $\mathcal{L}_b$ halts during the backpropagation. This causes the training process to bias towards the other loss function  $\mathcal{L}_d$\cite{Wang2021}. See \ref{Append:A} for complete details. Automatic update procedures for the Lagrange multiplier were developed to avoid biasing \cite{Wang2021,Van2022,Basir2022,Maddu2022}. However, most of these methods involve manual tuning of the hyperparameters for each frequency considered in the analysis. Hence, the latter approach is adopted throughout this work as it has only one loss function $\mathcal{L}_d$ to optimize.

The trial solution approach does not require any automatic update procedures and/or hyperparameter tuning. Therefore, it is more suitable to solve acoustic problems in the frequency domain. The following subsections demonstrate the application of this approach to three different problems in the area of plane wave duct acoustics. Note that throughout this article $p$ refers to the acoustic pressure and $u$ refers to the particle velocity. 

\subsection{Uniform duct} \label{Sec:2.1}
The acoustic pressure in a uniform duct can be obtained by solving the following one-dimensional (1-D) Helmholtz equation \cite{Morse1986}
\begin{equation}
    \left(\frac{d^2}{dx^2}+k^2\right)p(x)=0, \qquad x\in\left[x_1, x_2\right]
\end{equation}
where $k=\omega/c$ is the wavenumber, $\omega=2\pi f/c$ is the angular frequency, $c$ is the speed of sound, and $f$ is the frequency.

Let us assume the boundary conditions at $x_1$ and $x_2$ be
\begin{subequations}
\begin{align} 
p(x_1) &= p_1, \label{Eq:13a}\\
p(x_2) &= p_2. \label{Eq:13b}
\end{align}
\end{subequations}
If $\hat{p}(x;\theta)$ represents the neural network approximation for the acoustic pressure $p(x)$, the trial neural network $\hat{p}_t(x;\theta)$ that \emph{exactly} satisfy the chosen boundary conditions can be constructed as follows \cite{Lagaris1998}
\begin{equation}
    \hat{p}_t(x;\theta) = \phi_2p_1+\phi_1p_2+\phi_1\phi_2\hat{p}(x;\theta), \label{Eq:14}
\end{equation}
where $\phi_1$ and $\phi_2$ are the functions of the space variable $x$ which are to be constructed in such a way that
\begin{subequations}
\begin{align}
    \phi_1 &=  \begin{cases}
                0 &\text{at} \quad x=x_1,\\
                1 &\text{at} \quad x=x_2,
                \end{cases} \\
    \phi_2 &=  \begin{cases}
                1 &\text{at} \quad x=x_1,\\
                0 &\text{at} \quad x=x_2.
                \end{cases}
\end{align}
    
\end{subequations}

The functions that satisfy above mentioned rules can be constructed as follows
\begin{subequations}
    \begin{align}
    \phi_1 &= \frac{x-x_1}{x_2-x_1},\\
    \phi_2 &= \frac{x_2-x}{x_2-x_1}.
    \end{align}
\end{subequations}
It can be observed that the trial neural network $\hat{p}_t(x;\theta)$ has two parts. The first part contains only the boundary terms and does not contain any terms associated with the neural network $\hat{p}(x;\theta)$. On the contrary, the second part contains only the neural network and vanishes on the boundary. The first part ensures that the trial neural network satisfies the boundary conditions throughout the training process, whereas the second part contributes to the estimation of derivatives with respect to the space variable ($d/dx,\,d^2/dx^2$).

According to Eq.~(\ref{Eq:11}), the loss function which is to be optimized to predict the acoustic pressure can be written as follows
\begin{equation}
    \mathcal{L}(x;\theta) = \frac{1}{N}\sum_{i=1}^{N}\left\|\frac{d}{dx^{(i)}}\left(\frac{d}{dx^{(i)}}\hat{p}_t(x^{(i)};\theta)\right)+k^2\hat{p}_t(x^{(i)};\theta)\right\|^2_2. \label{Eq:17}
\end{equation}

\subsection{Duct with gradually varying cross-sectional area}
The plane wave acoustic field in a duct whose cross-sectional area changes gradually as a function of longitudinal position can be obtained by solving Webster equation given below \cite{Pillai2019}
\begin{equation}
    \left(\frac{d^2}{dx^2}+\frac{1}{S(x)}\frac{dS(x)}{dx}\frac{d}{dx}+k^2\right)p(x)=0, \qquad x\in\left[x_1, x_2\right] \label{Eq:18}
\end{equation}
where $S(x)$ is the cross-sectional area of the duct which varies as a function of $x\in\left[x_1, x_2\right]$.

The presence of additional term due to the area change makes the Webster equation differ significantly from the Helmholtz equation while solving it by the analytical means. However, in terms of the neural network formulation, there are no substantial differences in implementing solution procedures for both governing equations. The construction of the trial neural networks remains the same as in Eq.~(\ref{Eq:14}) provided the Webster equation is subject to similar boundary conditions as in Eqs.~(\ref{Eq:13a}) and (\ref{Eq:13b}). The only difference occurs in the calculation of the loss function $\mathcal{L}(x;\theta)$. The loss function in Eq.~(\ref{Eq:17}) should be replaced with 
\begin{multline}
     \mathcal{L}(x;\theta) = \frac{1}{N}\sum_{i=1}^{N}\left\|\frac{d}{dx^{(i)}}\left(\frac{d}{dx^{(i)}}\hat{p}_t(x^{(i)};\theta)\right)\right. \\
     +\left(\frac{1}{S(x^{(i)})}\frac{d}{dx^{(i)}}S(x^{(i)})\right)\left(\frac{d}{dx^{(i)}}\hat{p}_t(x^{(i)};\theta)\right) \\
     +\left.k^2\hat{p}_t(x^{(i)};\theta)\frac{}{}\right\|^2_2. 
\end{multline}

\subsection{Duct with narrow-tube effects} \label{Sec:2.3}
When the sound is propagating inside a duct, it will experience a dissipation through the visco-thermal effects. These effects are negligible when the sound is propagating through large cross-sectional ducts. However, when the sound is propagating through narrow ducts, these effects will become significant. It is for the same reason that these effects are also known as the \emph{narrow-tube effects}. Most of the models incorporate these effects in terms of the medium properties \cite{Allard2009}. 

The speed of sound and the density of air are considered to be real-valued and constant throughout the frequency range when analyzing sound propagation inside large cross-sectional ducts \cite{Munjal2014}. However, in the narrow ducts these properties are considered to be complex-valued, and functions of the frequency. Due to the narrow tube effects, the acoustic field inside the duct will become a complex-valued function \cite{Allard2009}. Performing an optimization process to generate complex-valued output is not a trivial task, as the loss function to be optimized has to be always a real-valued function. In this subsection, a methodology is presented which predicts the complex-valued acoustic pressure due to the narrow-tube effects in a uniform duct, by separating the real and imaginary-part of the governing equations as well as the boundary conditions \cite{Raissi2019}.

Let $c_w=c_R+jc_I$ and $k_w = k_R+jk_I = \omega/c_w$ be the complex-valued frequency-dependent speed of sound and wavenumber, respectively. Then, the complex-valued acoustic pressure $p_w=p_R+jp_I$ in a uniform duct can be obtained by solving
\begin{equation}
    \left(\frac{d^2}{dx^2}+k_w^2\right)p_w=0. \qquad x\in\left[x_1, x_2\right] \label{Eq:20}
\end{equation}
Here, the subscripts $R$ and $I$ represent the real and imaginary parts, respectively.  

Let the boundary conditions also be complex-valued
\begin{subequations}
\begin{align} 
p_w(x_1) &= p_R(x_1)+jp_I(x_1)=p_{1,R}+jp_{1,I}, \label{Eq:21a}\\
p_w(x_2) &= p_R(x_2)+jp_I(x_2)=p_{2,R}+jp_{2,I}. \label{Eq:21b}
\end{align}
\end{subequations}

\begin{figure}[h!]
	\includegraphics[scale=1]{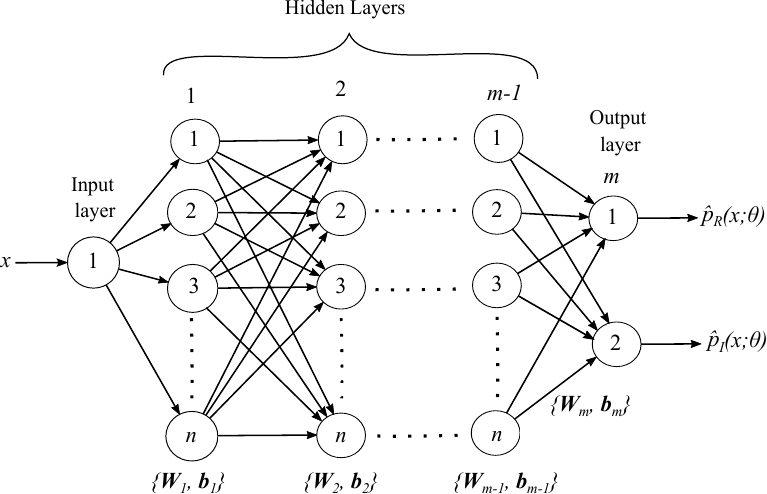}\centering
	\caption{\label{fig:2}{Schematic diagram of a feedforward neural network to predict complex-valued acoustic pressure.}}
	\end{figure}

Upon substituting $k_w$ and $p_w$, Eqs.~(\ref{Eq:20}), (\ref{Eq:21a}) and (\ref{Eq:21b}) can be arranged as two sets of governing equations. One set associated with the real-part and other set associated with the imaginary-part of the acoustic pressure as follows\\ \\
1) \underline{Equations associated with the real-part:} 
\begin{subequations}
\begin{align} 
\frac{d^2p_R}{dx^2}+\left(k_R^2-k_I^2\right)p_R-2k_Rk_Ip_I &= 0, \label{Eq:22a}\\
p_R(x_1) &=p_{1,R}, \\
p_R(x_2) &=p_{2,R}. \label{Eq:22c}
\end{align}
\end{subequations}
\noindent
2) \underline{Equations associated with the imaginary-part:} 
\begin{subequations}
\begin{align} 
\frac{d^2p_I}{dx^2}+\left(k_R^2-k_I^2\right)p_I+2k_Rk_Ip_R &= 0, \label{Eq:23a}\\
p_I(x_1) &=p_{1,I}, \\
p_I(x_2) &=p_{2,I}. \label{Eq:23c}
\end{align}
\end{subequations}
The corresponding trial neural network parts can be constructed as follows
\begin{align}
   \hat{p}_{t,R}(x;\theta) &= \phi_2p_{1,R}+\phi_1p_{2,R}+\phi_1\phi_2\hat{p}_R(x;\theta), \label{Eq:24} \\
   \hat{p}_{t,I}(x;\theta) &= \phi_2p_{1,I}+\phi_1p_{2,I}+\phi_1\phi_2\hat{p}_I(x;\theta), \label{Eq:25}
\end{align}
where $\hat{p}_R(x;\theta)$ and $\hat{p}_I(x;\theta)$ are the neural network approximations for the real and imaginary parts of the acoustic pressure, respectively. These can be predicted using the same neural network architecture as shown in Fig.~\ref{fig:1}, except that the output-layer should be replaced with a layer containing two neurons as shown in Fig.~\ref{fig:2}. One neuron is to predict the real-part and the other neuron is to predict the imaginary-part of the acoustic pressure. 

From Eqs.~(\ref{Eq:22a}) and (\ref{Eq:23a}), the loss function to be optimized can be written as follows
\begin{equation}
    \mathcal{L}(x;\theta) = \mathcal{L}_R(x;\theta) + \mathcal{L}_I(x;\theta), \label{Eq:26}
\end{equation}
where
\begin{multline}
     \mathcal{L}_R(x;\theta) = \frac{1}{N}\sum_{i=1}^{N}\left\|\frac{d}{dx^{(i)}}\left(\frac{d}{dx^{(i)}}\hat{p}_{t,R}(x^{(i)};\theta)\right)\right. \\
     +\left(k_R^2-k_I^2\right)\hat{p}_{t,R}(x^{(i)};\theta)-\left.2k_Rk_I\hat{p}_{t,I}(x^{(i)};\theta)\frac{}{}\right\|^2_2, 
\end{multline}
\begin{multline}
     \mathcal{L}_I(x;\theta) = \frac{1}{N}\sum_{i=1}^{N}\left\|\frac{d}{dx^{(i)}}\left(\frac{d}{dx^{(i)}}\hat{p}_{t,I}(x^{(i)};\theta)\right)\right. \\
     +\left(k_R^2-k_I^2\right)\hat{p}_{t,I}(x^{(i)};\theta)+\left.2k_Rk_I\hat{p}_{t,R}(x^{(i)};\theta)\frac{}{}\right\|^2_2. 
\end{multline}
It is important to note here that the Eqs.~(\ref{Eq:22a}) and (\ref{Eq:23a}) are coupled equations. They have to be solved simultaneously. In terms of the neural network formulation, this can be interpreted as finding the shared parameters which minimize the loss functions $\mathcal{L}_R$ and $\mathcal{L}_I$, simultaneously. This can be done by combining the loss functions as shown in Eq.~(\ref{Eq:26}) and performing the optimization procedure.

It is worth noting here that in Eq.~(\ref{Eq:26}), loss functions are added with equal weights in similar lines with the method discussed in \ref{Append:A}. It may be expected that it will introduce biasing behavior into the training process. The biasing occurs when one of the loss gradients vanishes during the backpropagation. This happens when we add the loss functions containing derivative terms of significantly different orders \cite{Maddu2022,Wang2021}. The derivative terms in the loss functions $\mathcal{L}_R$ and $\mathcal{L}_I$ in Eq.~(\ref{Eq:26}) are of the same order. Therefore, the chances are much lower to encounter a biasing problem.

\section{Results and Discussion}\label{Sec:3}
To predict the acoustic pressure, a neural network, as shown in Fig.~\ref{fig:1} is constructed with the architecture mentioned in Table~\ref{tab:1}. The information regarding the discretization of the domain and specifications of the optimizer used are also presented in it. The architecture is chosen in a similar way to that used by other researchers working in the area of physics-informed neural networks \cite{Raissi2019,Basir2022,Wang2021}. 
\begin{table}[h!]
    \centering
    \begin{tabular}{@{}clc@{}}
        \toprule
        Sl. No. & \hfil Parameter & Value \\
        \midrule
        1 & No. of layers ($m$) & 7 \\
        2 & No. of neurons in each hidden-layer ($n$) & 90 \\
        3 & Activation function ($\sigma$) & $\sin$ \\
        4 & No. of internal collocation points ($N_d$) & 14000 \\
        5 & No. of boundary collocation points ($N_b$) & 2 \\
        6 & Optimizer & L-BFGS \\
        7 & No. of iterations & 14000 \\
        8 & Optimal tolerance & 10$^{-3}$ \\
        \bottomrule
    \end{tabular}
    \caption{Neural network architecture and optimizer information}
    \label{tab:1}
\end{table}

The implementation is carried out in MATLAB (Version R2022b) with the help of the Deep Learning Toolbox$^{\text{\texttrademark}}$ and the Statistics and Machine Learning Toolbox$^{\text{\texttrademark}}$. It is to be noted here that during the implementation, the nonlinearity due to the activation function should be restricted to the hidden-layers only. The activation function of the output-layer should always be linear so that it does not limit the amplitude of the acoustic field to its nonlinear range ($\left[-1, 1\right]$).

\subsection{Acoustic pressure: Uniform duct} \label{Sec:3.1}
\begin{figure}[h!]
	\includegraphics[scale=0.8]{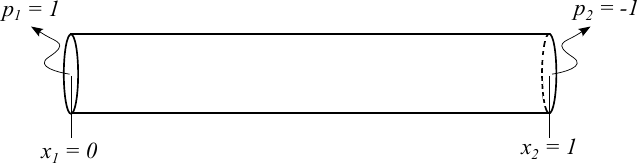}\centering
	\caption{\label{fig:3}{Schematic diagram of a uniform duct with boundary conditions.}} 
	\end{figure}

To predict the acoustic pressure in a uniform duct according to the formulation given in Section~\ref{Sec:2.1}, a duct of length 1 m as shown in Fig.~\ref{fig:3} is considered. The inlet and outlet boundary conditions (Eqs.~(\ref{Eq:13a}) and (\ref{Eq:13b})) are assumed to be $p_1=1$ and $p_2=-1$, respectively. 

\begin{figure}[h!]
	\includegraphics[scale=1.15]{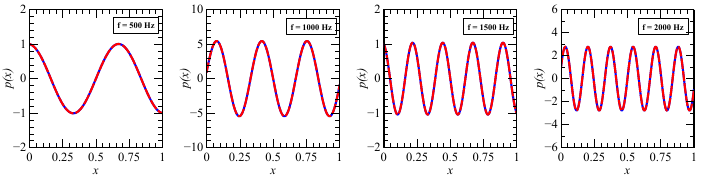}\centering
	\caption{\label{fig:4}{Acoustic pressure in a uniform duct:\protect\blueline true solution, \protect\redline predicted solution.}} 
	\end{figure}

The analysis is carried out between 500 Hz and 2000 Hz in steps of 500 Hz. The results are compared in Fig.~\ref{fig:4}. It can be observed that the acoustic pressure predicted by the neural network formulation using the trial solution approach (predicted solution) is in good agreement with that of the analytical method (true solution). Refer to \ref{Append:B} for the analytical solution.

The relative error between the two methods ($\delta p$) is calculated according to the following formulation and is used as a performance metric throughout the article.
\begin{equation}
    \delta p = \frac{\sqrt{\displaystyle\sum_{i=1}^{N_t} |\hat{p}_t(x^{(i)};\theta)-p(x^{(i)})|^2}}{\sqrt{\displaystyle\sum_{i=1}^{N_t}|p(x^{(i)})|^2}},
\end{equation}
where $N_t$ is the total number of linearly spaced test points in the domain $\left[0, 1\right]$, $p(x^{(i)})$ is the acoustic pressure calculated using the analytical method at $i$-th test point. In the current study, $N_t$ is taken as 500. The results for all four frequencies are tabulated in Table~\ref{tab:2}. It can be seen that the errors are almost negligible, which indicates that the neural network formulation is capable of learning the underlying physics from the governing equations at lower and higher frequencies with the trial solution approach.
\begin{table}[h!]
    \centering
    \begin{tabular}{@{}cc@{}}
        \toprule
        Frequency (Hz) & $\delta p$ \\
        \midrule
        500 &  8.5538$\times$10$^{-7}$\\
        1000 &  3.6359$\times$10$^{-5}$\\
        1500 &  7.7487$\times$10$^{-6}$\\
        2000 &  5.1628$\times$10$^{-5}$\\
        \bottomrule
    \end{tabular}
    \caption{Relative error between predicted and true solution for the uniform duct}
    \label{tab:2}
\end{table}

\subsection{Acoustic pressure: Duct with gradually varying cross-sectional area} \label{Sec:3.2}
\begin{figure}[h!]
	\includegraphics[scale=1]{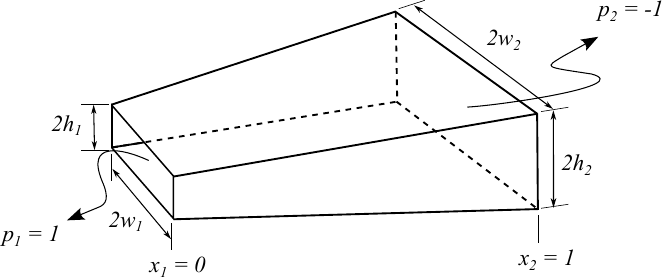}\centering
	\caption{\label{fig:5}{Schematic diagram of a gradually varying rectangular duct.}} 
	\end{figure}

Unlike the uniform duct, the estimation of the pressure distribution in a gradually varying cross-sectional duct by analytical means is not a trivial task. One has to adopt power series solutions \cite{Pillai2019,Yeddula2021} or a segmentation approach \cite{Gupta1995,Wang2011}. However, using neural networks, this can be achieved with the same effort as in the case of a uniform duct. To demonstrate this, a rectangular duct is considered whose cross-sectional area gradually varies as shown in Fig.~\ref{fig:5}. 
 
The length of the duct and the boundary conditions are assumed to be the same as those of the uniform duct. The cross-sectional area of the duct is assumed to be varying quadratically with respect to the position, i.e., $S(x)$ in Eq.~(\ref{Eq:18}) is assumed to be taken the form
\begin{equation}
    S(x) = S_0+S_1x+S_2x^2,
\end{equation}
where the coefficients $S_0$, $S_1$, and $S_2$, which can be found as \cite{Pillai2019}
\begin{align}
    S_0 &= 4h_1w_1, \\
    S_1 &= 4(h_1m_w+w_1m_h), \\
    S_2 &= 4m_hm_w.
\end{align}
Here, $m_h$ and $m_w$ are the slopes along height and width, respectively. These can be calculated as
\begin{align}
    m_h &= \frac{h_2-h_1}{x_2-x_1}, \\
    m_w &= \frac{w_2-w_1}{x_2-x_1}.
\end{align}
In the current study, the cross-sectional dimensions of the inlet are taken as $h_1=$ 10 mm and $w_1=$ 20 mm, and the dimensions of the outlet are taken as $h_2=3h_1$ and $w_2=3w_1$. 
	\begin{figure}[h!]
	\includegraphics[scale=1.15]{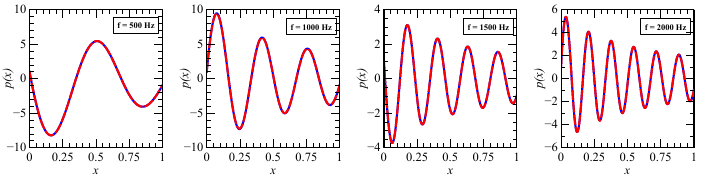}\centering
	\caption{\label{fig:6}{Acoustic pressure in a varying rectangular duct:\protect\blueline true solution, \protect\redline predicted solution.}}
	\end{figure}
 
Figure~\ref{fig:6} shows the comparison of the acoustic pressure obtained from the neural network formulation (predicted solution) against that obtained from the boundary value problem solver (\texttt{bvp4c}) in MATLAB (true solution). A pseudocode on the application of \texttt{bvp4c} to the current problem is given in \ref{Append:C}. It can be observed that the predicted solution agrees well with the true solution throughout the frequency range considered. The relative error between the two solutions is reported in Table~\ref{tab:3}. 
\begin{table}[h!]
    \centering
    \begin{tabular}{@{}cc@{}}
        \toprule
        Frequency (Hz) & $\delta p$ \\
        \midrule
        500 &  1.7972$\times$10$^{-4}$\\
        1000 &  6.584e$\times$10$^{-3}$\\
        1500 &  1.7072$\times$10$^{-4}$\\
        2000 &  2.6383$\times$10$^{-4}$\\
        \bottomrule
    \end{tabular}
    \caption{Relative error between predicted and true solution for gradually varying cross-sectional area duct}
    \label{tab:3}
\end{table}

It must be noted here that the peak value of the acoustic pressure in a uniform duct does not change with position, whereas in a gradually increasing area duct it decreases with respect to the position for a given boundary conditions. This is due to the damping characteristics introduced by the additional term (due to the area change) into the system.

\subsection{Acoustic pressure: Duct with narrow-tube effects} \label{Sec:3.3}
To predict the complex-valued acoustic pressure due to \emph{narrow-tube effects}, a uniform duct similar to the one in Section~\ref{Sec:3.1}, is considered with the same boundary conditions. The narrow-tube effects will become significant when the cross-sectional dimensions of the duct are smaller than the acoustic wavelength \cite{Allard2009}. Hence, the radius of the duct ($a$) is considered 0.5 mm. 

The expression for the frequency dependent complex-valued wavenumber ($k_w$) and the characteristic impedance ($z_w$) for narrow circular ducts is adopted from the literature and is given below \cite{Tijdeman1975,Stinson1991}
\begin{align}
    k_w &= k\times\sqrt{\frac{\gamma-(\gamma-1)\phi_h}{\phi_v}}, \\
    z_w &= z\times\sqrt{\frac{1}{\phi_v(\gamma-(\gamma-1)\phi_h)}},
\end{align}
where 
\begin{align}
    z &= \rho c, \\
    \phi_h &= -\frac{J_2(k_va)}{J_0(k_va)}, \\
    \phi_v &= -\frac{J_2(k_ha)}{J_0(k_ha)},
\end{align}
and 
\begin{align}
    k_v &= \sqrt{\frac{-j\omega\rho}{\mu}}, \\
    k_h &= \sqrt{\frac{-j\omega\rho c_p}{K}}.
\end{align}
Here, $\rho$ is the density, $\mu$ is the coefficient of dynamic viscosity, $c_p$ is the specific heat at constant pressure, $K$ is the thermal conductivity, $\gamma$ is the specific heat ratio, $k_v$ and $k_h$ are the viscous and thermal boundary wavenumbers, respectively. The validity of the model is subject to some assumptions which are stated in \ref{Append:D}. The values of the properties mentioned above for air under normal temperature and pressure (NTP) conditions are tabulated in Table~\ref{tab:4} \cite{Lemmon2000}. 
\begin{table}[h!]
    \centering
    \begin{tabular}{@{}clc@{}}
        \toprule
        Sl. No. & \hfil Parameter & Value \\
        \midrule
        1 & Density ($\rho$) & 1.225 kg/m$^3$ \\
        2 & Coefficient of dynamic viscosity ($\mu$) & 1.8$\times$10$^{-5}$ Pa.s\\
        3 & Specific heat at constant pressure ($c_p$) & 1007 J/kg.K \\
        4 & Thermal conductivity ($K$) & 0.02476 W/m.K \\
        5 & Specific heat ratio ($\gamma$) & 1.4 \\
        \bottomrule
    \end{tabular}
    \caption{Properties of air at 20$^{\circ}$C and 1 atm.}
    \label{tab:4}
\end{table}
	\begin{figure}[h!]
	\includegraphics[scale=1.15]{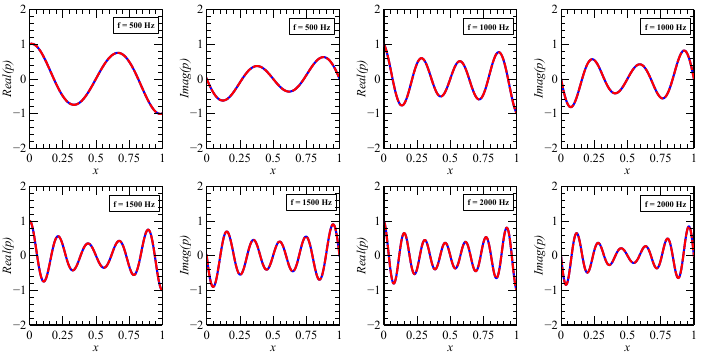}\centering
	\caption{\label{fig:7}{Acoustic pressure in a uniform duct with narrow tube effects:\protect\blueline true solution, \protect\redline predicted solution.}}
	\end{figure}
 
Figure~\ref{fig:7} shows both real and imaginary parts of the acoustic pressure obtained from the neural network (predicted solution) in comparison with the analytical solution (true solution, which can be obtained by replacing $k$ with $k_w$, $c$ with $c_w=\omega/k_w$, and $\rho$ with $\rho_w=z_w/c_w$ in \ref{Append:B}). The corresponding relative errors are reported in Table~\ref{tab:5}. It can be observed that the neural network is able to learn the underlying physics associated with both the real and imaginary parts of the equations simultaneously.
 \begin{table}[h!]
    \centering
    \begin{tabular}{@{}ccc@{}}
        \toprule
        Frequency (Hz) & $\delta p_{real}$ & $\delta p_{imag}$\\
        \midrule
        500 &  1.5225$\times$10$^{-6}$ & 2.6168$\times$10$^{-6}$\\
        1000 &  5.6833$\times$10$^{-6}$ & 8.6038$\times$10$^{-6}$\\
        1500 &  3.8647$\times$10$^{-5}$ & 3.1351$\times$10$^{-5}$\\
        2000 &  5.4766$\times$10$^{-4}$ & 2.4609$\times$10$^{-4}$\\
        \bottomrule
    \end{tabular}
    \caption{Relative error between predicted and true solution for the uniform duct with narrow-tube effects}
    \label{tab:5}
\end{table}

\subsection{Particle velocity estimation}
Besides acoustic pressure, it is essential to estimate the particle velocity to calculate the acoustic performance parameters of a system. This can be done using the momentum equation which gives the relationship between the acoustic pressure and particle velocity. 

\begin{figure}[h!]
	\includegraphics[scale=1.8]{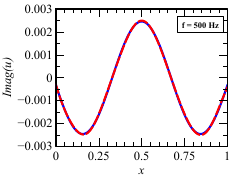}\centering
	\caption{\label{fig:8}{Particle velocity in a uniform duct without mean flow:\protect\blueline true solution, \protect\redline predicted solution.}}
	\end{figure}

In the absence of mean flow, the momentum equation results in \cite{Munjal2014}
\begin{equation}
    u = -\frac{1}{j\omega\rho}\frac{dp}{dx}. \label{Eq:43}
\end{equation}
If the expression of $p$ is known, it is trivial to estimate the particle velocity by analytical means. The neural network formulation offers a similar advantage. The network trained for the acoustic pressure $\hat{p}_{t}(x^{(i)};\theta)$ can be easily differentiated using automatic differentiation approach and the particle velocity can be estimated as follows
\begin{equation}
    \hat{u}_{t}(x^{(i)};\theta) = -\frac{1}{j\omega\rho}\frac{d}{dx^{(i)}}\hat{p}_{t}(x^{(i)};\theta).
\end{equation}
Note here that the subscript $t$ which is used to represent the trial solution approach is carry forwarded to the particle velocity as its estimation is based on the acoustic pressure obtained using the same approach. The results thus obtained for the uniform duct are compared against those of the analytical method for 500 Hz in Fig.~\ref{fig:8}. It can be observed that the predicted results are in good agreement with those of the true solution. Note here that since the acoustic pressure is a real-valued function, according to Eq.~(\ref{Eq:43}), the real-part of the particle velocity will be zero and the imaginary part is shown in Fig.~\ref{fig:8}.

It is trivial to estimate the particle velocity from the acoustic pressure in the absence of mean flow. However, when the mean flow is present, the momentum equation results in \cite{Li2017} 
\begin{equation}
    jku_w+M\frac{du_w}{dx}=-\frac{1}{\rho c}\frac{dp_w}{dx}, \label{Eq:45}
\end{equation}
where $M$ is the mean flow Mach number and the subscript $w$ denotes the complex-valued quantities. In this case, estimation of the particle velocity is not a straightforward process due to the presence of gradient terms associated with the particle velocity. Problems of this category can be solved using the \emph{transfer learning} technique \cite{Desai2021,Gao2022,Pellegrin2022,Xu2023}. According to this technique, the estimates of the acoustic pressure at discrete points $\hat{p}_{t}(x^{(i)};\theta)$ will be used to train another neural network $\hat{u}_{t}(x^{(i)};\Tilde{\theta})$ such that it satisfies
\begin{equation}
    jk\hat{u}_{t}(x^{(i)};\Tilde{\theta})+M\frac{d}{dx^{(i)}}\hat{u}_{t}(x^{(i)};\Tilde{\theta})=-\frac{1}{\rho c}\frac{d}{dx^{(i)}}\hat{p}_{t}(x^{(i)};{\theta}). \label{Eq:46}
\end{equation}
This results in another unconstrained optimization problem:
\begin{equation}
\min_{\Tilde{\theta}} \quad \mathcal{L}_u(x;\Tilde{\theta}), \quad x\in\left[x_1,x_2\right],
\end{equation}
where $\mathcal{L}_u$ is the loss function associated with the momentum equation. Since the presence of mean flow makes the acoustic pressure a complex-valued, the loss function $\mathcal{L}_u$ takes the form
\begin{equation}
    \mathcal{L}_u(x;\Tilde{\theta}) = \mathcal{L}_{u,R}(x;\Tilde{\theta}) + \mathcal{L}_{u,I}(x;\Tilde{\theta}), 
\end{equation}
where
\begin{align}
    \mathcal{L}_{u,R}(x;\Tilde{\theta}) &= \frac{1}{N}\sum_{i=1}^{N}\left\|M\frac{d}{dx^{(i)}}\hat{u}_{t,R}(x^{(i)};\Tilde{\theta})-k\hat{u}_{t,I}(x^{(i)};\Tilde{\theta})+\frac{1}{\rho c}\frac{d}{dx^{(i)}}\hat{p}_{t,R}(x^{(i)};\theta)\right\|^2_2,  \\
   \mathcal{L}_{u,I}(x;\Tilde{\theta}) &= \frac{1}{N}\sum_{i=1}^{N}\left\|M\frac{d}{dx^{(i)}}\hat{u}_{t,I}(x^{(i)};\Tilde{\theta})+k\hat{u}_{t,R}(x^{(i)};\Tilde{\theta})+\frac{1}{\rho c}\frac{d}{dx^{(i)}}\hat{p}_{t,I}(x^{(i)};\theta)\right\|^2_2.
\end{align}
Here, $\hat{u}_{t,R}$ and $\hat{u}_{t,I}$ are the real and imaginary parts of the particle velocity $\hat{u}_{t}$. The neural network formulation to predict the complex-valued acoustic pressure $\hat{p}_{t}(x^{(i)};\theta)$ is given in \ref{Append:E}. It is worth noting here that in the absence of mean flow, both the acoustic pressure and particle velocity share common network parameters, whereas in the presence of mean flow, they will have a different set of parameters. These parameters can be saved, and the networks can be reconstructed at later points of time to calculate other variables such as acoustic impedance, intensity, power, etc.

\begin{figure}[h!]
	\includegraphics[scale=1.6]{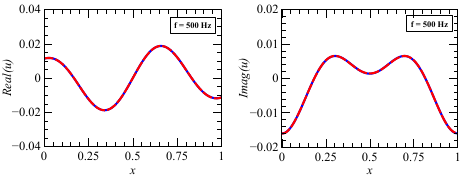}\centering
	\caption{\label{fig:9}{Particle velocity in a uniform duct with a mean flow of $M=$ 0.2:\protect\blueline true solution, \protect\redline predicted solution.}}
	\end{figure}

Figure~\ref{fig:9} shows the comparison of the real and imaginary parts of the particle velocity in a uniform duct at 500 Hz and $M=$ 0.2 against the analytical solution. Refer to \ref{Append:F} for the analytical solution. Here, the predictions of the neural network are obtained using the same architecture and optimization parameters mentioned in Table~\ref{tab:1}. The results reveal that the newly constructed network is able to predict the particle velocity from the existing network associated with the acoustic pressure successfully using the transfer learning technique.

One might be interested in finding the velocity potential ($\Phi$) instead of the particle velocity ($u$). In such cases, $u$ and $u_w$ in Eqs.~(\ref{Eq:43}) and (\ref{Eq:45}) can be replaced with the $d\Phi/dx$ and $d\Phi_w/dx$, and the velocity potential can be found from the acoustic pressure fields $p$ and $p_w$, respectively.

It should be noted here that the accuracy of the particle velocity (or the velocity potential) strongly depends on the accuracy of the acoustic pressure predictions. Therefore, one must be diligent in predicting the acoustic pressure. Any error that occurs in the prediction of it will propagate into the estimates of the particle velocity and subsequent variables.

In a typical neural network formulation, the training variables mentioned in Table~\ref{tab:1} will have a strong influence on the training of the network. Among those mentioned, the activation function ($\sigma$) and the number of collocation points ($N$) play a significant role in any physics-based neural network formulation \cite{Dung2023}. Choice of these variables based on the elementary knowledge on the output of the neural network significantly reduces the computational time and increases the accuracy. The following sections will present a study on the effect of these parameters while training a network to predict acoustic pressure in a uniform duct in the absence of mean flow. 

\subsection{Effect of activation function on the training process}
It is known that nonlinearity is introduced in the training process through the activation function \cite{Raissi2019,Wang2021,Basir2022}. In many physics-based neural network formulations, the hyperbolic tangent is used as an activation function \cite{Raissi2019,Basir2022}. However, it is not an ideal activation function for every problem. A suitable activation function for a particular neural network is purely problem specific. One should choose an activation function that better represents the output of the neural network. This will significantly reduce the computational time, and increase accuracy within chosen number of iterations. To demonstrate this, three activation functions, namely, \emph{hyperbolic tangent}, \emph{sine}, and \emph{cosine} are considered. 
	\begin{figure}[h!]
	\includegraphics[scale=1.8]{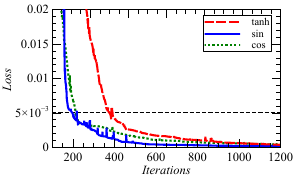}\centering
	\caption{\label{fig:10}{Loss function with respect to iterations for different activation functions at 500 Hz.}}
	\end{figure}

 Figure~\ref{fig:10} shows the comparison of the loss functions with respect to iterations for the three activation functions at 500 Hz. It can be seen that the functions \emph{sine} and \emph{cosine} are performing better as compared to \emph{hyperbolic tangent} function. For example, \emph{tanh} requires approximately 400 iterations to reach a chosen threshold loss value of 5$\times$10$^{-3}$, whereas \emph{sin} and \emph{cos} require half the iterations to reach the same value. It is due to the fact that both the activation function and the neural network output are sinusoidal in nature. 

 Table~\ref{tab:6} shows the relative errors for each activation function. The results indicate that both \emph{sin} and \emph{cos} activations offer more accurate results compared to \emph{tanh} within the chosen number of iterations (1200). Among \emph{sin} and \emph{cos}, the former activation function helps to reduce the loss function faster compared to the latter. Therefore, \emph{sin} is chosen as an activation function throughout this work.
\begin{table}[h!]
    \centering
    \begin{tabular}{@{}cc@{}}
        \toprule
        Activation function & $\delta p$ \\
        \midrule
        $\sin$ &  4.2369$\times$10$^{-4}$\\
        $\cos$ &  2.4916$\times$10$^{-4}$\\
        $\tanh$ &  17.0210$\times$10$^{-4}$\\
        \bottomrule
    \end{tabular}
    \caption{Relative errors with different activation functions at 500 Hz}
    \label{tab:6}
\end{table}
 
\subsection{Effect of number of data points on the training process}
\begin{figure}[h!]
	\includegraphics[scale=1.4]{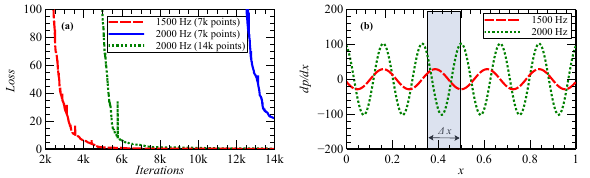}\centering
	\caption{\label{fig:11}{Effect of number of data points on the training process: (a) Loss function with respect to iterations for different number of data points, (b) gradient of acoustic pressure at low and high frequencies.}}
	\end{figure}

The choice of number of collocation points inside the domain significantly effects the training process, thence the accuracy of the solution. Unlike other problems, the required number of collocation points in acoustic problems depends on the frequency. The number of points appropriate for a particular frequency may not be adequate for another frequency. Figure~\ref{fig:11}a shows the loss function with respect to iterations for different frequencies and collocation points. It can be observed that 7000 points are sufficient to reduce the loss function at 1500 Hz to a sufficiently low value, whereas they are insufficient at 2000 Hz. The network requires double the number of points.

The accuracy of a neural network solution depends on how accurately the network is able to estimate the gradient of the variable being predicted with respect to the input variable. In the current work, estimation of $dp/dx$. Figure~\ref{fig:11}b shows the gradient of the acoustic pressure at 1500 Hz and 2000 Hz. It can be observed that the gradients at 2000 Hz are higher as compared to those at 1500 Hz. In other words, within the chosen subdomain, say $\Delta x\in\left[0, 1\right]$, the acoustic pressure variations are higher for 2000 Hz compared to 1500 Hz. Hence, the network needs finer spatial resolution at 2000 Hz as compared to 1500 Hz, to capture the variations in the acoustic pressure accurately. This requirement is not just attributed to high-frequency analysis. Accurate pressure predictions at near-resonance/natural frequency also require higher spatial resolution. 
	\begin{figure}[h!]
	\includegraphics[scale=1.4]{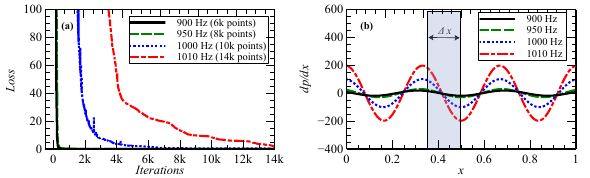}\centering
	\caption{\label{fig:12}{Effect of number of data points on the training process at near-resonance frequency: (a) Loss function with respect to iterations as frequency approaches to resonance frequency, (b) gradient of acoustic pressure as frequency approaches to resonance frequency.}}
	\end{figure}

For the uniform duct configuration considered in the current study, resonance occurs when $\sin(kL)=$ 0, i.e., at the frequencies
\begin{equation}
    f_n = \frac{nc}{L},
\end{equation}
where $f_n$ is the natural frequency, $L$ is the length of the duct and $n=$ 1, 2, 3,..., so on. Figure~\ref{fig:12} demonstrates the effect of the number of collocation points on the training process near one of these resonance frequencies (1020 Hz). Figure~\ref{fig:12}a shows the variation of the loss function as the frequency approaches the resonance frequency from the low frequency side. For 900 Hz and 950 Hz, which are away from 1020 Hz by 120 Hz and 70 Hz, respectively, 6000 - 8000 collocation points are enough for the network to reduce the loss function to a sufficiently low value within 1000 iterations. As the frequency approaches the resonance frequency (1000 Hz and 1010 Hz), the network needs almost double the number of points to reduce the loss function to a sufficiently low value. Figure~\ref{fig:12}b shows the corresponding pressure gradients. From the results, it is evident that as the analysis frequency approaches one of the resonance/natural frequencies, the acoustic pressure gradients drastically increases and the network needs finer spatial resolution to capture these gradients to satisfy the governing differential equation. 
\newpage \noindent
\section{Conclusion}\label{Sec:4}
Predicting acoustic pressure in the frequency domain using the neural network method is a challenging task due to the \emph{vanishing gradient problem}. In addition, prediction of a complex-valued acoustic field is not a straightforward process, as existing algorithms perform optimization only on the real-valued functions. These two challenges are addressed in the current paper with examples from plane wave acoustic theory. The vanishing gradient problem is circumvented by constructing a trial neural network that always satisfies the boundary conditions prior to training. The complex-valued acoustic field is predicted by splitting the governing equations and the boundary conditions into real and imaginary parts and training the network to identify shared parameters that satisfy both parts simultaneously. 

Research findings reveal that a single neural network architecture is capable of predicting acoustic field in a uniform duct with visco-thermal and convective mean flow effects with a maximum error of the order $\mathcal{O}(10^{-4})$. The same network architecture is able to predict the acoustic field in a gradually varying cross-sectional area duct with a maximum error of $\mathcal{O}(10^{-3})$. The study further reveals the following
\begin{enumerate}[label=(\alph*)]
    \item Through the transfer learning, particle velocity and other parameters can be estimated easily.
    \item In the case of frequency domain acoustics, the number of collocation points required to train a network successfully is inversely proportional to the iterations. In other words, a network with fewer number of input collocation points requires more iterations than the network with adequate number of input collocation points.
    \item The network needs finer spatial resolution at higher frequencies and near-resonance frequencies to accurately capture the spatial variation of the acoustic field. 
\end{enumerate}

These observations are not just limited to plane wave duct acoustics. They are applicable to problems in higher-dimensional spaces such as in the prediction of the acoustic field in 2D and 3D geometries. The ability of a single neural network architecture to predict the acoustic field in a wide range of problems will lead to the development of meshless acoustic solvers, which can be deployed on open platforms in the near future. 

\section*{Data availability}
Data will be made available on request.

\section*{Acknowledgments}
The authors acknowledge the support received from the Department of Science and Technology, and Science and Engineering Research Board (SERB), Government of India towards this research.

\appendix
\section{Illustration of bias towards one of the loss functions while training a neural network} \label{Append:A}
Let us consider a problem of solving one-dimensional Helmholtz equation to predict the acoustic pressure in a uniform duct of length 1 m:
\begin{equation}
    \left(\frac{d^2}{dx^2}+k^2\right)p(x)=0, \qquad x\in\left[0, 1\right]
\end{equation}
subjected to the boundary conditions 
\begin{subequations}
\begin{align} 
p(0) &= 1, \\
p(1) &= -1. 
\end{align}
\end{subequations}
According to Raissi et al. \cite{Raissi2019}, the loss function to be minimized to obtain the neural network approximation for the acoustic pressure, $\hat{p}(x;\theta)$, can be written as follows
\begin{equation}
    \mathcal{L}(x;\theta) = \mathcal{L}_d(x;\theta) + \mathcal{L}_b(x;\theta), 
\end{equation}
i.e., $\lambda_j=$ 1, $j=$ 1, 2 in Eq.~(\ref{Eq:9}), and
\begin{align}
     \mathcal{L}_d(x;\theta) &= \frac{1}{N_d}\sum_{i=1}^{N_d}\left\|\frac{d}{dx^{(i)}}\left(\frac{d}{dx^{(i)}}\hat{p}(x^{(i)};\theta)\right)+k^2\hat{p}(x^{(i)};\theta)\right\|^2_2, \\
     \mathcal{L}_b(x;\theta) &= \frac{1}{N_b}\sum_{i=1}^{N_b}\left\|\hat{p}(x^{(i)};\theta)-p(x^{(i)})\right\|^2_2.
\end{align}
To minimize $\mathcal{L}$, a feedforward neural network with 5 hidden layers and 90 neurons in each hidden layer is constructed. The domain is divided into 14000 random collocation points apart from the two boundary points. The optimization is performed using \emph{L-BFGS} optimizer with \emph{sin} activation function. A total of 14000 iterations were performed with an optimal tolerance of 10$^{-3}$. 

\begin{figure}[h!]
	\includegraphics[scale=1.8]{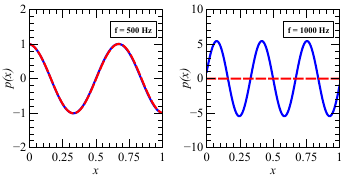}\centering
	\caption{\label{fig:A1}{Acoustic pressure in a uniform duct with $\lambda_j=$ 1:\protect\blueline true solution, \protect\redline predicted solution.}}
	\end{figure}

Figure~\ref{fig:A1} shows the comparison of the acoustic pressure obtained from the predictions of the neural network (predicted solution) and the analytical method (true solution). It can be observed that at lower frequencies (500 Hz), the results from both methods are in good agreement with each other. As the frequency increases, the neural network is unable to learn the underlying physics from the governing equation, which ultimately results in bad correlation. The reason for this behavior can best be understood by observing the individual loss functions and their gradients.
 
\begin{figure}[h!]
	\includegraphics[scale=2]{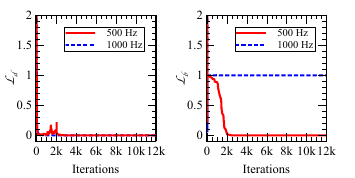}\centering
	\caption{\label{fig:A2}{Loss functions with respect to iterations for different frequencies}}
	\end{figure}

Figure~\ref{fig:A2} shows the comparison of individual loss functions with respect to iterations at different frequencies. It can be seen that the loss functions associated with the differential equation ($\mathcal{L}_d$) reduce to zero for both frequencies. However, the loss function associated with the boundary conditions ($\mathcal{L}_b$) does not reduce and becomes stagnant at a particular value with a high frequency. This indicates that the training process is biased towards $\mathcal{L}_d$ at higher frequencies.  
 
\begin{figure}[h!]
	\includegraphics[scale=2]{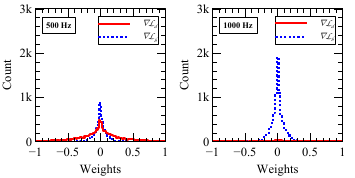}\centering
	\caption{\label{fig:A3}{Histograms of loss gradients}}
	\end{figure}

Figure~\ref{fig:A3} shows the histograms of the gradients calculated at the last hidden layer of the network in the last iteration for the two frequencies. It can be observed that at higher frequencies (1000 Hz), the gradients of $\mathcal{L}_b$ are concentrated at zero, while the gradient of $\mathcal{L}_d$ are evenly distributed. Since most of the gradients of $\mathcal{L}_b$ are in the neighborhood of zero, they will vanish during backpropagation. This causes the training process to bias towards minimizing $\mathcal{L}_d$, leaving $\mathcal{L}_b$, at higher frequencies.

Algorithms are proposed that update the Lagrange multiplier such that the vanishing gradient problem is circumvented \cite{Basir2022,Van2022,Wang2021,Maddu2022}. However, most of these algorithms require manual tuning of the hyperparameter(s) for each frequency. For instance, Eq.~(\ref{Eq:A6}) shows a popular $\lambda_j$ update rule in which $\lambda_j$ at iteration $\tau+1$ is updated based on the value of $\lambda_j$ and $\hat{\lambda}_j$ at iteration $\tau$ \cite{Wang2021,Maddu2022}:  
\begin{equation}
    \lambda_{j}(\tau+1) = \alpha\lambda_j(\tau)+(1-\alpha)\hat{\lambda}_j(\tau). \label{Eq:A6}
\end{equation}
Here, $\hat{\lambda}_i(\tau)$ is updated based on the statistical parameters calculated from the gradients of individual loss functions. $\alpha$ is a user-defined hyperparameter that decides the contribution of $\lambda_j$ from the current iteration towards the next iteration. This hyperparameter requires manual tuning for each frequency considered in the analysis.

\section{Analytical solution for the acoustic pressure distribution in a uniform duct} \label{Append:B}
The acoustic pressure in a uniform duct can be obtained by solving 1-D Helmholtz equation \cite{Morse1986}
\begin{equation}
    \left(\frac{d^2}{dx^2}+k^2\right)p(x)=0, \qquad x\in [x_1,\,x_2], \label{Eq:B1}
\end{equation}
subjected to the boundary conditions
\begin{equation}
    p(x_1)=p_1; \qquad p(x_2)=p_2.
\end{equation}
Since the governing differential equation is homogeneous, it can be treated as a eigenvalue problem and can be solved by assuming the pressure distribution of the form
\begin{equation}
    p(x) = Ce^{\lambda x}, \label{Eq:B3}
\end{equation}
where $\lambda$ is an eigenvalue. Upon substituting Eq.~(\ref{Eq:B3}) in Eq.~(\ref{Eq:B1}) yields the solution 
\begin{equation}
    p(x)=C_1\cos(kx)+C_2\sin(kx), \label{Eq:B4}
\end{equation}
where $C_1$ and $C_2$ are constants which can be calculated from the boundary conditions. 

Upon substituting the boundary conditions in Eq.~(\ref{Eq:B4}), the constants can be obtained as 
\begin{align}
    C_1 &= \frac{p_1\sin(kx_2)-p_2\sin(kx_1)}{\sin(k(x_2-x_1))}, \\
    C_2 &= \frac{p_1\cos(kx_2)-p_2\cos(kx_1)}{\sin(k(x_1-x_2))}.
\end{align}

The particle velocity $u(x)$ can be calculated using the momentum equation (Eq.~(\ref{Eq:43})) as follows
\begin{equation}
    u(x) = -\frac{j}{\rho c}\left[C_1\sin(kx)-C_2\cos(kx)\right].
\end{equation}
\newpage

\section{Pseudocode to calculate pressure distribution in a gradually varying cross-sectional area duct using  \texttt{bvp4c}} \label{Append:C} 
\begin{breakablealgorithm}
  \begin{algorithmic}[1]
  \Require $x_1, x_2, p_1, p_2, S_0, S_1, S_2, k, N_t$
  \State $x \gets \texttt{linspace}(x_1,x_2,N_t)$
  \item[] 
  \Statex{{\textcolor{blue}{\footnotesize\texttt{/*  Function to predict the acoustic pressure  */}}}}
    \Function{\texttt{pressure}}{} 
        \State $ sol \gets \texttt{bvp4c\small(ODEFCN,BCFCN,INITSOL)}$
        \State $p \gets sol(:,1)$ \Comment{\textcolor{blue}{\footnotesize\texttt{1$^{st}$ column:= $p$; 2$^{nd}$ column:= $p^{\prime}$}}}
    \EndFunction
    \item[] 
    \Statex{{\textcolor{blue}{\footnotesize\texttt{/*  Function to calculate $p$ and $p^{\prime}$  */}}}}
    \Function{\texttt{odefcn}}{$x, \mathbf{y}, S_0, S_1, S_2, k$}
        \State $ S \gets S_0+S_1x+S_2x^2$
        \State $ S^{\prime} \gets S_1+2S_2x$
        \State $\mathbf{y} \gets \left[y_1; \quad y_2\right]$ \Comment{\textcolor{blue}{\footnotesize\texttt{$y_1$ and $y_2$ represents $p$ and $p^{\prime}$, respectively.}}}
        \State $\mathbf{y}^{\prime} \gets \left[y_2; \quad -S^{\prime}y_2/S-k^2y_1\right]$
    \EndFunction
    \item[] 
    \Statex{{\textcolor{blue}{\footnotesize\texttt{/*  Function to calculate difference in boundary values  */}}}}
    \Function{\texttt{bcfcn}}{$\mathbf{y}, x_1, x_2, p_1, p_2$}
        \State $ \mathbf{y_b} \gets \left[y_1(x_1)-p_1; \quad y_1(x_2)-p_2\right]$
    \EndFunction
    \item[] 
    \Statex{{\textcolor{blue}{\footnotesize\texttt{/*  Function to initialize $p$ and $p^{\prime}$  */}}}}
    \Function{\texttt{initsol}}{$x, \mathbf{y}$}
        \State $ \mathbf{y_0} \gets \left[\texttt{sin}(x); \quad \texttt{cos}(x)\right]$
    \EndFunction
  \end{algorithmic}
\end{breakablealgorithm}

\newpage

\section{Assumptions for the validity of complex-valued wavenumber $k_w$} \label{Append:D}
The validity of the complex-valued wavenumber adopted from the literature is subjected to the following assumptions \cite{Allard2009,Tijdeman1975,Stinson1991}
\begin{itemize}
    \item Acoustic wavelength should be much larger than the boundary layer thickness, i.e., viscous and thermal wavenumbers must be much larger than the acoustic wavenumber
    \begin{equation}
        \frac{k_v}{k} >> 1; \quad \frac{k_h}{k} >> 1.
    \end{equation}
     \item The cross-section of the duct must be much smaller than the acoustic wavelength so that the plane wave theory is valid
     \item The cross-section of the duct must be constant or at most slowly varying in the direction of sound propagation
     \item The length of the duct in the propagation direction must be larger than the boundary layer thickness, i.e.,
     \begin{equation}
         L >> \delta_v; \quad L >> \delta_h,
     \end{equation}
     where 
     \begin{equation}
         \delta_v = \sqrt{\frac{2\mu}{\omega\rho}}; \quad \delta_h = \sqrt{\frac{2K}{\omega\rho c_p}},
     \end{equation}
     are the viscous and thermal boundary layer thicknesses, respectively
\end{itemize}
These assumptions can be easily verified for the medium properties and duct configuration considered in Section~\ref{Sec:3.3}.

\section{Neural network formulation to predict the complex-valued acoustic pressure in a uniform duct in the presence of mean flow} \label{Append:E}
The acoustic pressure in a uniform duct in the presence of mean flow can be obtained by solving \cite{Li2017,Yeddula2021}
\begin{equation}
    \left[(1-M^2)\frac{d^2}{dx^2}-2jMk\frac{d}{dx}+k^2\right]p_w(x)=0, \qquad x\in\left[x_1, x_2\right], \label{Eq:E1}
\end{equation}
where $M=U/c$ is the mean flow Mach number, $U$ is the mean flow velocity, and $j=\sqrt{-1}$.

The presence of an additional gradient term due to the mean flow makes the acoustic pressure a complex-valued function. Following the approach in Section~\ref{Sec:2.3}, Eq.~(\ref{Eq:E1}) along with the boundary conditions Eqs.~(\ref{Eq:21a}) and (\ref{Eq:21b}), produces two sets of equations. One set for the real-part and the other set for the imaginary-part of the acoustic pressure. The corresponding optimization problem can be constructed as 
\begin{equation}
\min_{\theta} \quad \mathcal{L}_R(x;\theta)+\mathcal{L}_I(x;\theta), \quad x\in\left[x_1, x_2\right], 
\end{equation}
where $\mathcal{L}_R$ and $\mathcal{L}_I$ are the loss functions associated with the real and imaginary parts, respectively. They can be calculated as
\begin{multline}
        \mathcal{L}_R(x;\theta) = \frac{1}{N}\sum_{i=1}^{N}\left\|(1-M^2)\frac{d}{dx^{(i)}}\left(\frac{d}{dx^{(i)}}\hat{p}_{t,R}(x^{(i)};\theta)\right)\right. \\
        +\left.2Mk\frac{d}{dx^{(i)}}\hat{p}_{t,I}(x^{(i)};\theta)+k^2\hat{p}_{t,R}(x^{(i)};\theta)\right\|^2_2, 
\end{multline}

\begin{multline}
    \mathcal{L}_I(x;\theta) = \frac{1}{N}\sum_{i=1}^{N}\left\|(1-M^2)\frac{d}{dx^{(i)}}\left(\frac{d}{dx^{(i)}}\hat{p}_{t,I}(x^{(i)};\theta)\right)\right. \\
        -\left.2Mk\frac{d}{dx^{(i)}}\hat{p}_{t,R}(x^{(i)};\theta)+k^2\hat{p}_{t,I}(x^{(i)};\theta)\right\|^2_2.
\end{multline}
Here, $\hat{p}_{t,R}$ and $\hat{p}_{t,I}$ are the trial solutions associated with the real and imaginary parts of the acoustic pressure, respectively, and can be constructed as in Eqs.~(\ref{Eq:24}) and (\ref{Eq:25}).

\section{Analytical solution for the complex-valued acoustic pressure in a uniform duct in the presence of mean flow} \label{Append:F}
Since Eq.~(\ref{Eq:E1}) is an ordinary differential equation with constant coefficients, by following the similar procedure as in \ref{Append:B}, the general solution can be obtained as
\begin{equation}
    p_w(x)=C_{w,1}e^{-jk_c^+x}+C_{w,2}e^{jk_c^-x}, \label{Eq:F1}
\end{equation}
where $k_c^+=k/(1+M)$ and $k_c^-=k/(1-M)$ are the convective wavenumbers. The constants $C_{w,1}$ and $C_{w,2}$ can be obtained from the boundary conditions (Eqs.~(\ref{Eq:21a}) and (\ref{Eq:21b}))  as follows
\begin{align}
    C_{w,1} &= \frac{p_1e^{jk_c^-x_2}-p_2e^{jk_c^+x_1}}{e^{j(k_c^-x_2-k_c^+x_1)}-e^{j(k_c^-x_1-k_c^+x_2)}}, \\
    C_{w,2} &= \frac{p_1e^{-jk_c^+x_2}-p_2e^{-jk_c^+x_1}}{e^{j(k_c^-x_1-k_c^+x_2)}-e^{j(k_c^-x_2-k_c^+x_1)}}.
\end{align}
Here, the boundary values $p_1$  and $p_2$ can be purely real or complex valued. 

The particle velocity $u_w$ can be assumed to take the form as
\begin{equation}
    u_w(x)=D_{w,1}e^{-jk_c^+x}+D_{w,2}e^{jk_c^-x}, 
\end{equation}
and the constants $D_{w,1}$ and $D_{w,2}$ can be calculated from the momentum equation (Eq.~(\ref{Eq:45})) as \cite{Munjal2014}
\begin{align}
    D_{w,1} &= \frac{C_{w,1}}{\rho c}, \\
    D_{w,2} &= -\frac{C_{w,2}}{\rho c}.
\end{align}

\bibliographystyle{elsarticle-num} 
\bibliography{references}

\end{document}